\documentclass[journal=jacsat,manuscript=article]{achemso}

\usepackage[version=3]{mhchem} 

\author{Emanuela Margapoti}
\email{emanuela.margapoti@wsi.tum.de}
\affiliation{Walter Schottky Institute - ZNN, Physik Department and NIM, Technische Universit\"{a}t M\"{u}nchen, Am Coulombwall 4, Garching, Germany}
\phone{+49 (0)289 11366}
\fax{+49 (0)289 12704}

\author{Philipp Strobel}
\affiliation{Walter Schottky Institute - ZNN, Physik Department and NIM, Technische Universit\"{a}t M\"{u}nchen, Am Coulombwall 4, Garching, Germany}

\author{Mahmoud M. Asmar}
\altaffiliation{Department of Physics and Astronomy and Nanoscale and Quantum Phenomena Institute, Ohio University, Athens, Ohio 45701-2979, USA}
\affiliation{Dahlem Center for Complex Quantum Systems and Fachbereich Physik, Freie Universit\"at Berlin, 14195 Berlin, Germany}

\author{Max Seifert}
\affiliation{Walter Schottky Institute - ZNN, Physik Department and NIM, Technische Universit\"{a}t M\"{u}nchen, Am Coulombwall 4, Garching, Germany}

\author{Juan Li}
\affiliation{Physik Department E20, Technische Universität München, James-Franck-St. 1, Garching,Germany}

\author{Matthias Sachsenhauser}
\affiliation{Walter Schottky Institute - ZNN, Physik Department and NIM, Technische Universit\"{a}t M\"{u}nchen, Am Coulombwall 4, Garching, Germany}

\author{\"{O}zlem Ceylan}
\affiliation{Walter Schottky Institute - ZNN, Physik Department and NIM, Technische Universit\"{a}t M\"{u}nchen, Am Coulombwall 4, Garching, Germany}

\author{Carlos-Andres Palma}
\affiliation{Physik Department E20, Technische Universität München, James-Franck-St. 1, Garching,Germany}

\author{Johannes V. Barth}
\affiliation{Physik Department E20, Technische Universität München, James-Franck-St. 1, Garching,Germany}

\author{Jose A. Garrido}
\altaffiliation{Walter Schottky Institute - ZNN, Physik Department and NIM, Technische Universit\"{a}t M\"{u}nchen, Am Coulombwall 4, Garching, Germany}

\author{Anna Cattani-Scholz}
\altaffiliation{Walter Schottky Institute - ZNN, Physik Department and NIM, Technische Universit\"{a}t M\"{u}nchen, Am Coulombwall 4, Garching, Germany}

\author{Sergio E. Ulloa}
\altaffiliation{Department of Physics and Astronomy and Nanoscale and Quantum Phenomena Institute, Ohio University, Athens, Ohio 45701-2979, USA}
\affiliation{Dahlem Center for Complex Quantum Systems and Fachbereich Physik, Freie Universit\"at Berlin, 14195 Berlin, Germany}

\author{Jonathan J. Finley}
\affiliation{Walter Schottky Institute - ZNN, Physik Department and NIM, Technische Universit\"{a}t M\"{u}nchen, Am Coulombwall 4, Garching, Germany}

\title[An \textsf{achemso} demo]
  {Emergence of photoswitchable states in a graphene-azobenzene-Au platform}

\abbreviations{IR,NMR,UV}

\begin{document}

\begin{abstract}
The perfect transmission of charge carriers through potential barriers in graphene (Klein tunneling) is a direct consequence of the Dirac equation that governs the low-energy carrier dynamics. As a result, localized states do not exist in unpatterned graphene, but quasi-bound states \emph{can} occur for potentials with closed integrable dynamics. Here, we report the observation of resonance states in photo-switchable self-assembled molecular(SAM)-graphene hybrid. Conductive AFM measurements performed at room temperature reveal strong current resonances, the strength of which can be reversibly gated \textit{on-} and \textit{off-} by optically switching the molecular conformation of the mSAM. Comparisons of the voltage separation between current resonances ($\sim 70$--$120$ mV) with solutions of the Dirac equation indicate that the radius of the gating potential is $\sim 7 \pm 2$ nm with a strength $\geq 0.5$ eV. Our results and methods might provide a route toward \emph{optically programmable} carrier dynamics and transport in graphene nano-materials.
\end{abstract}


Predictions of resonant states in quantum Dirac billiards\cite{Berry87} have spurred intense experimental interest exploring, for example, the quasi-bound states in nano-ribbons\cite{Silvestrov07} and the relationship they have with the classical integrability of the dynamics, as dictated by the geometry\cite{Bardarson09,Wurm11,Zhao-Tan12,Ponomarenko08,Libisch09}. A variety of different approaches were explored to create bound or quasi-bound states in graphene, including the definition of n-p-n regions using local gates\cite{Young09}. Direct etching or oxidation of graphene was also used to produce nano-ribbons\cite{Jiao09} or quantum dot like structures\cite{Ponomarenko08,Rozhkova11} known to have unique transport properties. Other approaches, including adatom deposition or ion bombardment through masks have also been used to produce well-defined regions where the carriers experience strong confinement\cite{Archanjo12}. Whilst these diverse approaches to obtain carrier confinement are very successful they are all system specific and, moreover, are inflexible in the sense that they require changes in the fabrication processes in order to modify the confinement strength and geometry.
In the present manuscript, we demonstrate tuning of the local potential to \emph{optically} gate the potential landscape in graphene-hybrid systems via self-assembled monolayer, employing a switchable method.

\begin{figure}
\centering
\includegraphics[scale=0.5]{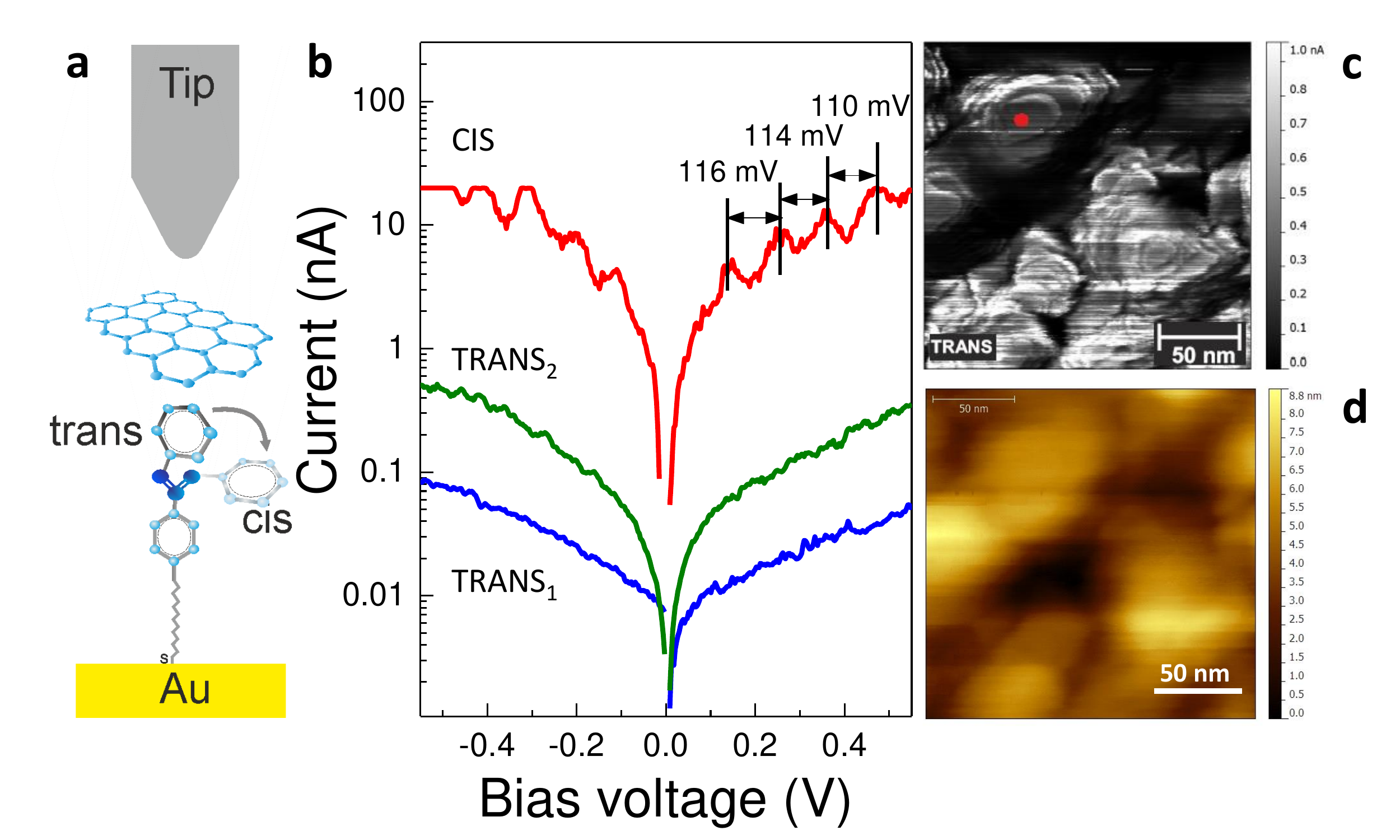}
\caption{\small{\textbf {Samples and measurement configurations}}. \textbf{a}, Schematic representation of c-AFM measurement on the graphene/mSAM/Au hybrid with the azobenzene molecule in {\em trans} (elongated form) and {\em cis} (bent) configurations. \textbf{b}, Typical I-V characteristics in {\em trans}-configuration before illumination (blue trace), in {\em cis}-configurations after switching with UV-light (red trace) and after switching back to the \emph{trans}-configuration (green trace) following white light illumination. \textbf{c}, Typical current topography over a $200\times200$ nm area at a fixed bias voltage of 10 mV with the mSAM in the {\em trans}-configuration. The red spot indicate the position of the AFM-cantilever when the I-V was performed. The current compliance used for all measurements was set to $20$ nA.\textbf{d}, topographical image of \textbf{b}.}
\end{figure}

The samples investigated were fabricated using an Au-surface functionalised with a mixed self-assembled monolayer (mSAM), consisting of an azobenzene derivative (4-(1-mercapto-6-hexyloxy)-azobenzene) and spacer-molecules (6-mercapto-1-hexanol) such as to prevent steric hindrance of the photo-mediated molecular conformation switching. Further detail of the final sample fabrication, graphene/mSAM/Au heterostructure, is reported in the supplementary information (SI). Azobenzene molecules are known to be photochromic, being able to reversibly switch their conformation from the thermodynamically stable \emph{trans}-configuration to the metastable \emph{cis}-configuration upon illumination with UV light\cite{Bandara12, Qune08, Crivillers13}. This optically induced photo-isomerisation is depicted schematically in Fig. 1a and results in a reduction of the length of the molecule from $1.90$ nm to $1.45$ nm and a significant change of the dipole moment\cite{Klajn10,Seo13}. We performed contact angle measurements to confirm that the mSAM molecular configuration is indeed fully switched by illumination at $365$ nm with a power density of $850$ $\mu$Wcm$^{-2}$ for about 40 $\pm$ 10 minutes (see SI - Fig. S2b). The complete switching back to $trans$-configuration was obtained illuminating the sample with white light for about two hours.

Results obtained from two different classes of samples are investigated here; a {\em reference} sample consisting of a mSAM on Au substrate without graphene (see SI - Fig. S4) and two different types of {\em active} samples in which a monolayer of mechanically exfoliated graphene\cite{Novoselov05} was positioned on top of the mSAM to form the graphene/mSAM/Au hybrid. The only difference between the two $active$ samples is the use of two different substrates, i.e. the Au deposited on glass substrate and the Au(111) deposited on mica substrate. The results using Au(111) are reported in the SI. In the remainder of the paper we refer to these two types of samples as the reference (graphene/Au) and active samples (graphene/mSAM/Au), respectively. The transport properties of all samples and the topography of the local conductivity were characterized at room temperature in ambient condition, using c-AFM in contact mode in three configurations; (i) before illumination, in which the azobenzene layer is in the \emph{trans}-configuration, (ii) after illumination with UV-light ($365$ nm with $850$ $\mu$Wcm$^{-2}$ for 30 mins) to induce a metastable switch of the molecular conformation to the \emph{cis}-configuration, and (iii) after photo-switching the mSAM back to the {\em trans}-configuration using white light illumination for 40 min. C-AFM measurements were performed using a commercial scanning probe microscope employing a Pt-Ir coated Si cantilever with a spring constant of ~2 N/m. Resonance frequency of the cantilever was approximately 70 kHz. The c-AFM measurements were executed in contact mode allowing measuring the current density through the metal tip to study the transport properties of both the reference (Fig. S4 in SI) and active samples (Fig. 1-2) with the mSAM in either the \emph{cis}- and \emph{trans}- configurations.

The reference sample (Fig.S4 in the SI) shows the expected behavior upon switching the molecular conformation, that is, the current increasing by a factor of 10 when the mSAM is switched into the {\em cis} configuration. This behavior is expected, arising from the reduction of the molecular length and the subsequent lowering of the tunneling barrier length, as reported by other groups\cite{Mativetsky08,Kronemeijer08}.

Differently, in Fig.1b typical local I-V transport measurements performed on one of the active sample are presented and characterized using the Au on glass as a substrate. Here, a dramatic and reversible change of the conductivity upon switching the mSAM from the \emph{trans}- to the \emph{cis}- configuration is in the red and blue traces. In strong contrast to the \emph{trans}- configuration, in the \emph{cis}-configuration, pronounced peaks are observed in the I-V characteristics, with a typical voltage gap between the peaks of $\delta V_{gap}\sim 100 \pm 20$ meV. Such peaks in the local transport characteristics are \emph{only} observed in the active sample, and only when the mSAM is switched into the \emph{cis}-configuration, vanishing when the sample is switched back to the \emph{trans}-configuration (see green curve in Fig. 1b). Notably, the increase of current from \emph{trans}- to \emph{cis}-configuration is a factor of 100 stronger. The current map shown in Fig.1c and recorded at $+10$ mV (in the \emph{trans}-configuration) reveals terraces with rings of high current, delineating the grains of the underlying Au-electrode when deposited on glass substrate (Fig.1d).

\begin{figure}
\centering
\includegraphics[scale=0.7]{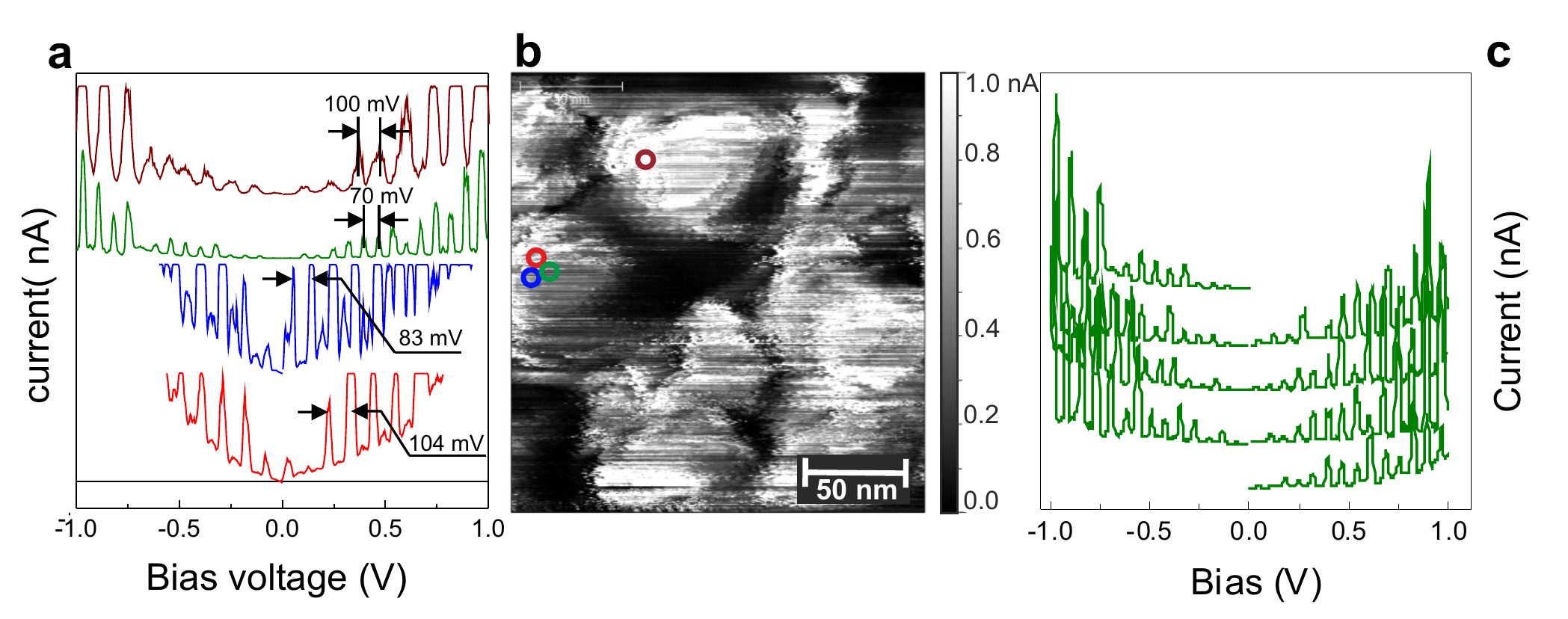}
\caption{\small{\textbf{Current resonances}}. \textbf{a}, I-V characteristics measured on several positions of the sample. Colour of the I-V traces corresponds to the position on the current map recorded at fixed bias of 10 mV when the molecules are in {\em cis}-configuration, \textbf{b}. The switching of the molecules has been carried out by illuminating the sample with 360 nm lamp over a 30 min time scan. \textbf{c}, Typical I-V traces recorded consecutively in up-down sweeps over an entire 2 sec time. Each trace is vertically shifted for clarity; the arrow indicates bias sweep direction. The resonance peaks are slightly shifted by few millivolts with respect to the previous cycle.}
\end{figure}

For the graphene/mSAM/Au sample, the peak to valley ratio of the current resonances observed in the \emph{cis} I-V characteristics was found to be dependent on the exact position of the Pt-Ir tip on the surface. This is shown in Fig. 2a that compares several different I-V traces, plotted on a linear scale and recorded at different locations on the sample surface, as indicated by the circles on the current map image (Fig. 2b), and recorded after UV exposure. Pronounced current resonances are observed in all the I-V characteristics in this \emph{cis}-configuration with peak spacing of about $70$ $\pm$ $20$ mV, superimposed on an approximately linear background, increasing with bias voltage. Measurements were performed on different locations on the grains in the graphene/mSAM/Au-glass hybrid, universally showing the presence of pronounced current oscillations only when the mSAM is in the {\em cis}-configuration. It is worth to mention that the peak spacing dependence from the tip localization demonstrate that our observation is only to be inferred to the graphene/mSAM/Au hybrid and not to the graphene-single-molecule interaction.
Note that the amplitude of the current peaks is clipped to a maximum of $20$ nA due to the current compliance of the amplifier used in the c-AFM. Upon executing consecutive up and down voltage sweeps we observed weak, but clear hysteretic effects as shown in Fig. 3c. Clear strong resonances are observed in the current, with peaks being rigidly shifted by $\left|\Delta V\right|\sim3-12$ mV between successive sweeps, with respect to the previous trace. This observation is indicative of voltage induced fluctuations in the local potential in the graphene/mSAM/Au hybrid multilayers.

\begin{figure}[h!]
\centering
\includegraphics[scale=0.5]{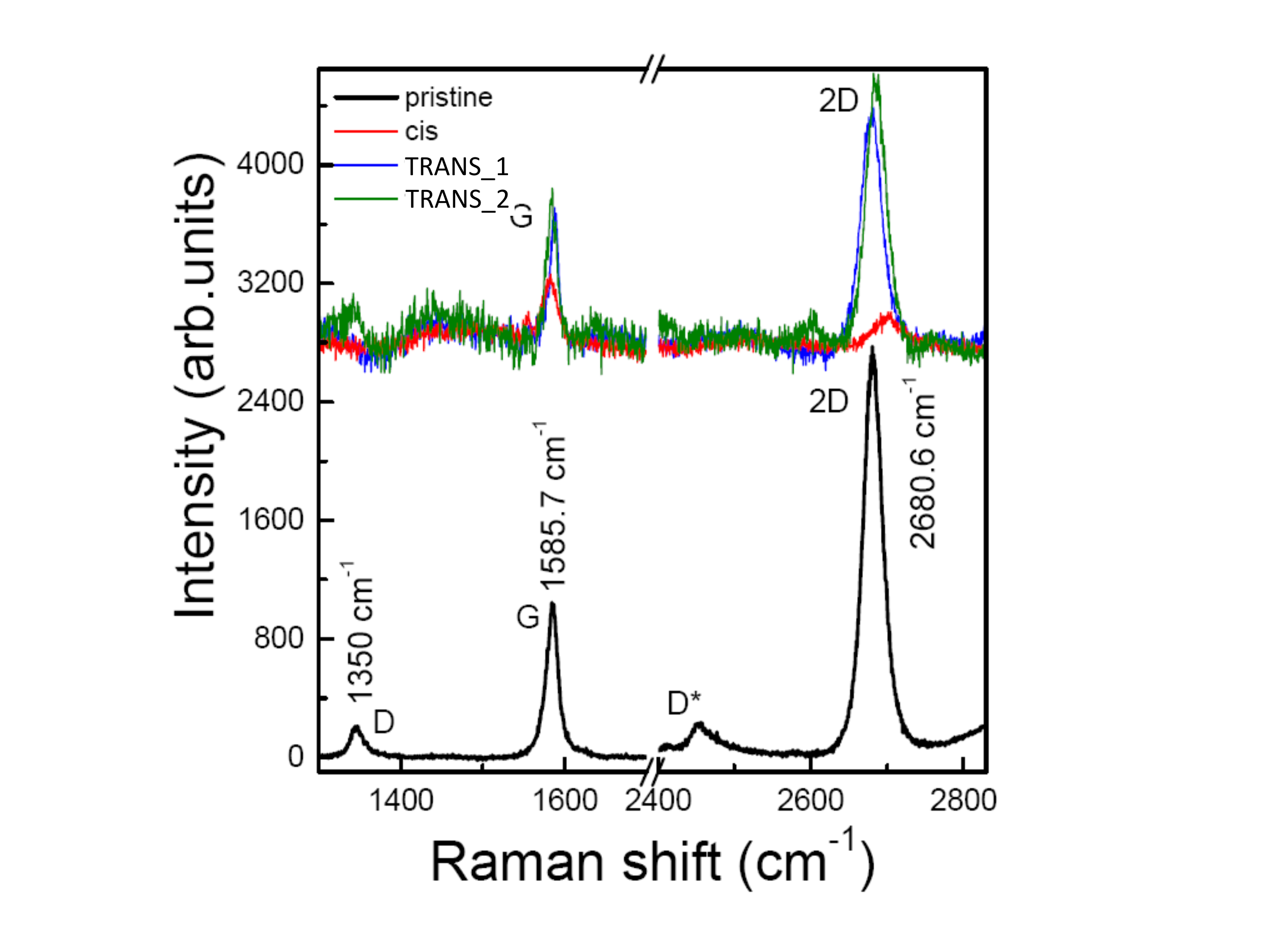}
\caption{\small{\textbf{Raman spectroscopy on graphene/$SiO_2$ and Graphene/Azobenzene/Au hybrid system}}. \textbf{a}, Raman spectrum recorded from a pristine graphene on $SiO_2$ (dark curve) when excited with 514.5 nm cw laser. The top spectra refers to the graphene when transferred on mSAM on Au substrate. The blue and green spectra refer to the case when the azobenzene molecules are in $trans$, while the red spectrum was recorded just after 30 min of UV exposure, when the azobenzene molecules are in $cis$. The switching back of the molecules was obtained by irradiating the sample with white light for 40 min. }
\end{figure}

To gain more insight into the modulation the graphene electronic landscape caused by photoswitching of the mSAM, we performed Raman spectroscopy (Fig. 3). In particular, one can use the Raman data as a finger print of the defect density (D-band), the in-plane carbon $sp{^2}$ vibrational mode (G-band) and stacking orders (2D-bands). It has also been demonstrated by several groups\cite{MALARD09,DAS08,Pisana07} that the frequency of the G and 2D Raman peaks, as well as their linewidth and the ratio of their intensity (I${_{2D}}$/I${_G}$), are dependent on the position of the Fermi level relative to the Dirac point. \cite{PEIMYOO12}

Figure 3 compares typical Raman data recorded from pristine graphene and from the graphene/ mSAM/Au hybrid with the mSAM switched into \emph{cis}- and \emph{trans}- configurations. Starting from pristine graphene (black spectrum) the ratio I${_{2D}}/I{_G}$ which is usually above two is lowered to a value $1 < I{_{2D}}/I{_G} < 2$ for the graphene/mSAM/Au hybrid when the molecules are in {\em trans} (blue spectrum) indicative of doping effect.
Upon switching the mSAM to the {\em cis}-configuration (red spectrum), the intensity ratio I${_{2D}}/I{_G}$ is even below one, indicative of a stronger shift of the Fermi level from the Dirac point and significant interaction between the mSAM and the graphene layer as suggested above. Finally, upon switching the mSAM back to {\em trans}-configuration, after illumination with white light for 90 min, we observe a near full recovery of the observed Raman signal (green spectrum). The drop in I${_{2D}}/I{_G}$ recorded in $cis$ suggests that the proximal mSAM results in a shift or gating of the Fermi level induced by the azobenzene molecules already in {\em trans} and an even stronger perturbation in {\em cis}-configuration. The shift of the Raman peaks provides evidence of doping from the mSAM in {\em trans}-configuration, since both the G and 2D bands are blue-shifted with respect to the reference sample by about 4--8 cm$^{-1}$.
It should be emphasized here the role of the 2D peak and its suppression when the molecules are in $cis$-configurations. The 2D peak is the overtone of the the D-peak and it is a peak that originates from a two scattering process and therefore momentum conservation is always satisfied by two phonons; then no defects are required for their activation. From this we understand that a strong perturbation is required in order to modify the 2D peak. Indeed, in presence of a high doping a dramatic change of the 2D peak was reported showing a ratio I${_{2D}}/I{_G}$ that is alway below one for each excitation energy (see SI in Ferrari et al. \cite{Ferrari13}). Even more conclusive is the attenuation of the 2D peak observed by Cancado et al.\cite{Cancado11} when the point-like defects in graphene reach nanometric size. Therefore, there is strong evidence toward the resonance oscillations measured in our system graphene/mSAM/Au originating from quasi-bound states formed in localized regions formed in the graphene/mSAM system as the Raman spectroscopy demonstrate.

\begin{figure}
\centering
\includegraphics[scale=0.6]{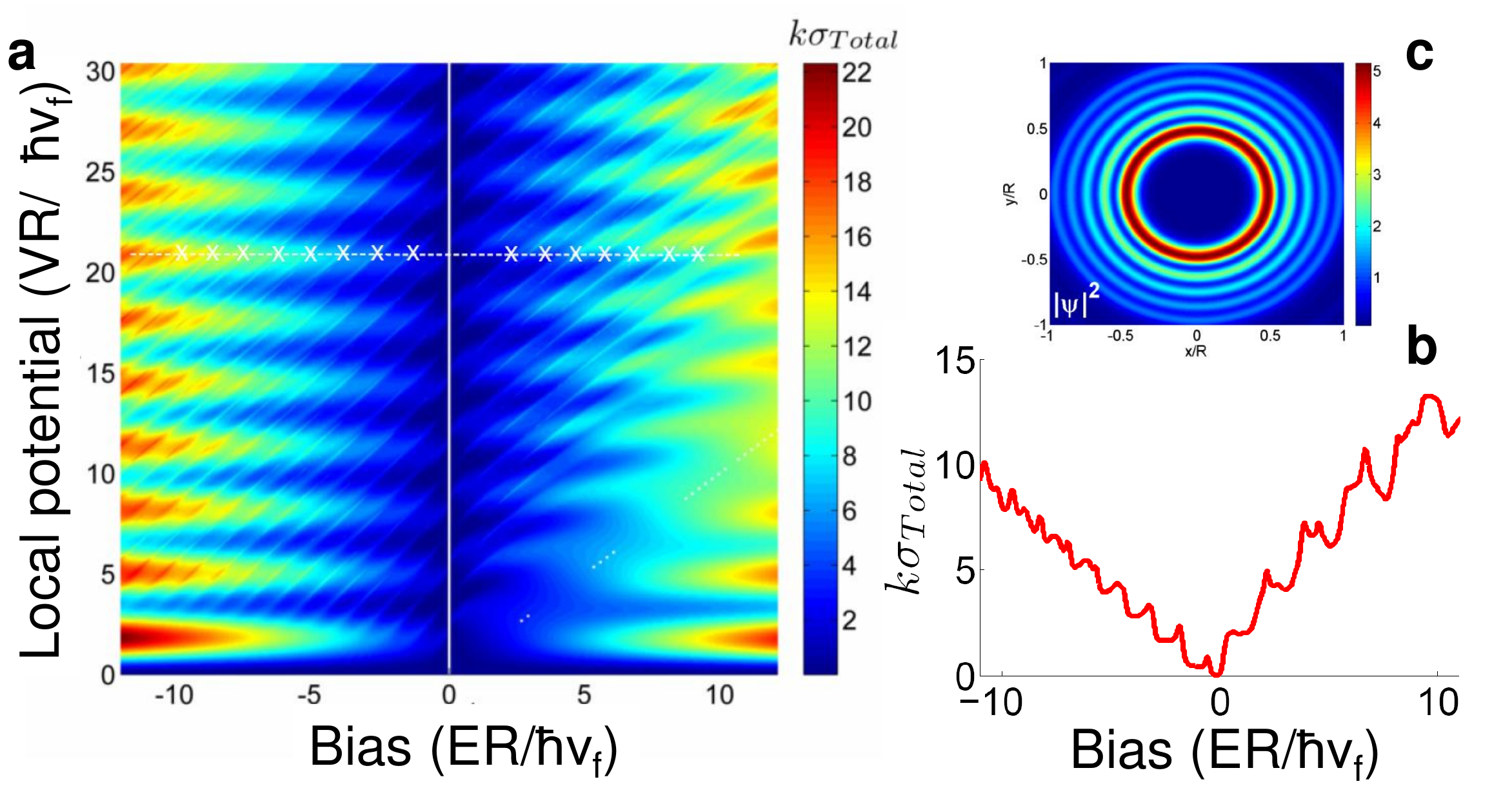}
\caption{\small{\textbf{Shape resonances in a circular region}}.
\textbf{a}, Map of total cross section, $\sigma_{Total}$, for
circular gated region of radius $R$, as function of gate potential
$V$ and electron energy $E$ in units of $\hbar v_F/R$
($\sigma_{Total}$ scaled by $k=ER/\hbar v_F$).  Nearly periodic
peaks in $E$ at constant $V$ are shape resonances with consecutive
peaks characterized by $\Delta E \simeq \hbar v_F/R$. \textbf{b},
Line trace of total cross section as a function of $ER/\hbar v_{F}
\propto V_{Bias}$, for $VR/\hbar v_F = 22.1$ with thermal broadening
corresponding to room temperature (see SI).
The peaks are nearly equally spaced after the thermal broadening
(nearly periodic resonances correspond to the white crosses in panel
a). \textbf{c}, Spatial amplitude map of a typical resonance at
$ER/\hbar v_F = -10.2$ for $VR/\hbar v_F = 22.1$ (leftmost white
cross in panel a and leftmost peak in panel b), resonant peaks in
the cross section result in spatial amplitudes increasingly bounded
to the perimeter of the circular gate region as $|E|$ increases,
similarly to well-known whispering gallery modes.}
\end{figure}

In order to simulate the effect of local gating, the 2D Dirac equation with a potential perturbation was solved for a circular region. The circular region is an approximation of the areas formed in the graphene/mSAM hybrid system. The potential perturbation can be described over a finite-sized circular region, $H_V = V\vartheta(R-r)$, where $V$ represents the potential strength of the gated region, and $\vartheta$ is the Heaviside function determining the radius $R$ of the region. Assuming the potential to be smooth such that it does not cause intervalley scattering, one can treat the problem at each Dirac point independently\cite{Katsnelson09,Asmar13}. The eigenvalue problem becomes ${\hbar v_F}(\sigma_xp_x + \sigma_yp_y)\stackrel{\rightarrow}{\psi} = {(E-V\vartheta(R-r))}\stackrel{\rightarrow}{\psi}$, where $\stackrel{\rightarrow}{\psi}$ is a spinor in the pseudospin basis, and $\sigma$ are Pauli matrices. The cylindrical symmetry of the problem allows one to use a partial wave decomposition, so that the phase shifts gained by each angular momentum channel during the scattering process are obtained analytically. The phase shifts can be directly related to differential, total and transport cross sections, which reveal different aspects of the scattering processes, including the formation of quasi-bound states in the scattering region.

Figure 4 shows a characteristic local potential versus bias map of
the total cross section for this problem. High amplitude peaks in
this map signal the presence of resonances in the scattering as
marked by the white crosses (at $VR/\hbar v_F = 22.1$), indicating
the formation of periodic quasi-bound states in the gated region
with finite life time characterized by the width of the peak. Here,
the thermal broadening $\simeq 25$meV appears to be the dominant
factor controlling the resonance line widths (see SI), suggesting that the injected electrons propagate
ballistically over the characteristic length scales explored in our
experiment. The appearance of such `shape resonances' is a direct
consequence of interference effects of waves scattering on the
potential profile. The quasi-bound states can be seen as circulating
inside the gated region, avoiding Klein tunneling for a finite time,
given their scattering from the potential boundary at non-normal
angles of incidence(see Fig. 4b). For high energies, the energy
difference $\Delta E$ between two consecutive peaks allows us to
determine the characteristic size of the gated regions, as $\Delta E
\simeq \hbar v_{F}/R$ (see SI). A resonance
spacing of ~100 meV, as seen in experiments, yields ~7 nm for the
size of the gated region induced by the cis-azobenzene mSAM.
\cite{Mahmoud14}  Regions with other spacings, such as $\Delta E \simeq 70$ meV
in Fig.\ 2,
yield a size ~9.5 nm, reflecting the different potential landscapes defined by
the azobenzene gating. 

It is important to notice that an irregular geometry would result in shorter lifetimes for the shape resonances, becoming increasingly short-lived at higher energies (bias), but that their equal spacing in energy would remain in the high quantum index limit, corresponding to the semiclassical regime.

Three explanations for the observed resonances in the quasi-bound state framework are suggested. One is that the local electrostatic potential is patterned by the mSAM, whereby 6-mercapto-1-hexanol can readily deprotonate, creating electrostatic traps. Azobenzenes embedded in the mSAM when in $cis$-configuration would allow maximum interaction of the underlying 6-mercapto-1-hexanol with the graphene and turn on the formation quasi-bound states. A second explanation centers around the local electrostatic potential being tuned by work function changes induced by the mSAM.\cite{Qune08,Crivillers13} A third possibility, concerns the roughness of the substrate, where the quasi-bound states would be dictated by terraces and step edges between grains, as observed in the current images in Figs. 1c,d. In this view, the $cis$-configurations will allow closer contact with the underlying substrate causing the graphene to abruptly fold into morphological traps.

In summary, we have presented experimental evidence for the direct observation of resonance oscillations in the current density of a graphene-mSAM-Au hybrid, formed by reversible switching of the azobenzene molecules. The functionalization of the substrate underneath graphene enables the reversible modification of the electrical and quantum properties of the Dirac fermions. Such variation is obtained by using aromatic photochromic molecules that change their configuration upon illumination with UV light. We demonstrate that when the molecules are in the {\em cis} configuration, the mSAM locally gates the nearby graphene layer, resulting in resonance oscillations of the current density as recorded by c-AFM spectroscopy. This hybrid assembly allows for possible {\em in situ} modification of scattering potentials on graphene using a convenient method - simply illuminating to change the conformation of the proximal mSAM in a controllable fashion.

\textbf{Supporting Information Available}

Experimental section, and conductive AFM measurements of pristine graphene and reference samples were included. This material is available free of charge via the Internet at http://pubs.acs.org. 

\textbf{Corresponding author}

*E-mail: emanuela.margapoti@wsi.tum.de

\textbf{Notes}

The authors declare no competing financial interest.

\begin{acknowledgement}

We are grateful to Gerhard Abstreiter, Per-Lennart Adelt and V. L\'opez-Richard for fruitful discussions. We thank the DFG for financial support via the Nanosytems Initiative Munich. M.M.A. and S.E.U. supported in part by NSF-CIAM/MNW grant DMR-1108285. They are also grateful for the welcoming environment at the Dahlem Center and the support of the A. von Humboldt Foundation.

\end{acknowledgement}

\end{document}